\documentstyle[12pt,epsf]{article}
\makeatletter
\@addtoreset{equation}{section}
\makeatother

\begin{document}

\begin{titlepage}

\def\thefootnote{\fnsymbol{footnote}}
\begin{flushleft}
\setlength{\baselineskip}{13pt}
ITP-UH-12/95	\hfill	March 1995\\
cond-mat/9504103
\end{flushleft}
\vspace*{\fill}

\begin{center}
{\Large Thermodynamics of an integrable model for electrons\\
        with correlated hopping} \\
\vfill
\vspace{1.5 em}
{\sc Gerald Bed\"urftig}\footnote{e-mail: {\tt bed@itp.uni-hannover.de}}
{\sc and Holger Frahm}\footnote{e-mail: {\tt frahm@itp.uni-hannover.de}}\\
{\sl Institut f\"ur Theoretische Physik, Universit\"at Hannover \\
D-30167~Hannover, Germany}\\
\vfill
ABSTRACT
\end{center}

\begin{quote}
\setlength{\baselineskip}{13pt}
A new supersymmetric model for electrons with generalized hopping terms
and Hubbard interaction on a one-dimensional lattice is solved by
means of the Bethe Ansatz. We investigate the phase diagram of this model
by studying the ground state and excitations of the model as a function of
the interaction parameter, electronic density and magnetization. Using
arguments from conformal field theory we can study the critical exponents
describing the asymptotic behaviour of correlation functions at long
distances.
\end{quote}

\vfill
PACS-numbers:
71.27.+a~\	
75.10.Lp~\	
05.70.Jk~\	

\vfill
\setcounter{footnote}{0}
\end{titlepage}

\section{Introduction}
In recent years studies of one dimensional models of electronic systems
have been a primary source to gain understanding of correlation effects in
low dimensional systems. In particular the growing number of exactly
soluble models such as the {\em Bethe Ansatz} integrable Hubbard and
supersymmetric $t$--$J$ models and their extensions have provided new
insights into ground state properties of these systems \cite{liwu:68}%
\nocite{suth:75,schl:87,eks:92a}---\cite{barx:93}.

Different sources of interaction have been studied in these models: Apart
from the influence of the on-site Coulomb repulsion (which is the main
physical motivation leading to the Hubbard model) and the antiferromagnetic
coupling of electrons leading to spin fluctuations (as present in the
$t$--$J$ model) the kinetic energy can been modified to include interaction
effects. Such {\em bond-charge} repulsion terms reflecting the dependence
of nearest neighbour hopping amplitudes on the occupation of sites affected
were first discussed in Ref.~\cite{hirsch:89b}. There have been extensive
studies of the relevance of such additional interaction terms for example
in their relation to the possibility of superconductivity based on
electronic correlations (see e.g.\ \cite{hima:90,boks:95}).
Furthermore, several exact solutions for one-dimensional models of this type
have been found (see e.g.\ \cite{eks:92a,barx:93,aral:94}).

In this paper we consider a new integrable model containing generalized
hopping integrals that has been found recently \cite{bglz:95,befr:95c}. The
Hamiltonian is given as
\begin{eqnarray}
{\cal H} = &-& \sum_i \sum_{\sigma=\uparrow\downarrow}
	\left(c_{i\sigma}^\dagger c_{i+1\sigma} +h.c.\right)\
	\Bigl(t_0 - X \left( n_{i,-\sigma}+n_{i+1,-\sigma}\right)
	  +\bar{X}\, n_{i,-\sigma} n_{i+1,-\sigma} \Bigr)
\nonumber \\
   &-&t_3\ \sum_i \left( c_{i+1,\uparrow}^\dagger c_{i+1,\downarrow}^\dagger
	      c_{i\downarrow} c_{i\uparrow} + h.c.\right)
   + U\ \sum_i n_{i\uparrow} n_{i\downarrow}
\label{eq:Hamil} \\
   &-&\mu\ \sum_i (n_{i\uparrow}+n_{i\downarrow})
   - {h\over2}\ \sum_i (n_{i\uparrow}-n_{i\downarrow})\ .
\nonumber
\end{eqnarray}
In addition to the usual single particle hopping amplitude $t_0$ and the
on-site Coulomb integral $U$ it contains the bond-charge interaction $X$,
an additional coupling $\bar{X}$ correlating hopping amplitudes with the
local occupation, and a pair hopping term with amplitude $t_3$. In addition,
the Hamiltonian contains coupling to a chemical potential $\mu$ and
magnetic field $h$ controlling particle density and magnetization of the
system, respectively.

For later convenience we introduce a different parametrization of the
hopping integrals by $X=t_0-t_1$, $\bar{X}=t_0-2t_1+t_2$. Studying the
two particle scattering matrix ${\cal S}$ one finds two possible choices
of the parameters $t_j$ and $U$ where ${\cal S}$ satisfies a Yang Baxter
equation resulting in candidates for models (\ref{eq:Hamil}) that might
be integrable by means of the Bethe Ansatz: first, choosing $t_0=t_1=t_2$
(which implies $X=\bar{X}=0$) and $t_3=0$ the Hamiltonian reduces to the
well known Hubbard model \cite{liwu:68}. Another family of such models
arises for\footnote{%
The special cases $t_2=t_0/2$ and $t_2=2t_0$ have been discussed before
in \cite{karn:94}.}
\begin{equation}
	{1\over2}\ U=-t_3=\pm(t_0-t_2)\ne 0\ , \qquad
	(t_1)^2=t_0 t_2
\label{eq:solvpt}
\end{equation}
In an independent approach, the integrability of the model
(\ref{eq:Hamil}) with (\ref{eq:solvpt}) has been proven in the framework
of the Quantum Inverse Scattering method where the Hamiltonian has been
derived from a solution of the Quantum Yang-Baxter equation invariant under
a four dimensional representation of $gl(2|1)$ \cite{bglz:95} which is the
symmetry underlying the (Bethe Ansatz soluble) supersymmetric $t$--$J$ model.

Our paper is organized as follows: In the following section we shall
discuss the symmetries of the Hamiltonian (\ref{eq:Hamil}) at the
integrable point (\ref{eq:solvpt}). It turns out that there are two
physically different regions to be studied corresponding to $t_0>t_2$ and
$t_0<t_2$ (or positive and negative $U$), respectively. In Section~3 the
Bethe Ansatz equations determining the spectrum of the model are derived.
In Section~4 ground state properties and the spectrum of low-lying excitations
at temperature $T=0$ are determined and in Section~5 we shall study finite
size corrections of the spectrum to discuss the asymptotic behaviour of
correlation functions. In the Appendix we discuss the completeness of the
solutions obtained from these equations for small systems.

\section{Symmetries}
Owing to various symmetries of the Hamiltonian (\ref{eq:Hamil}) only the
upper sign in the relation (\ref{eq:solvpt}) with positive $t_0$ and $t_2$
has to be studied:

To see this, we first note that the sign of $t_1$ is not fixed by the
conditions (\ref{eq:solvpt}). In fact, the unitary transformation
\begin{equation}
   c_{i\sigma} \rightarrow c_{i\sigma}(1-2 n_{i,-\sigma})
\end{equation}
has the only effect of changing $t_1 \rightarrow -t_1$.

A particle--hole transformation performs a mapping between $t_0,t_2>0$ and
$t_0,t_2<0$ (as a consequence of (\ref{eq:solvpt}) $t_0$ and $t_2$ necessarily
have the same sign!):
\begin{equation}
   T_1:\quad c_{i\sigma} \rightarrow c_{i\sigma}^\dagger\ ,
       \quad \sigma=\uparrow, \downarrow\ .
\end{equation}
Applying this transform to the Hamiltonian we obtain (the irrelevant change
of sign in $t_1$ is suppressed)
\begin{eqnarray}
  &&{\cal H}(t_0,t_2,U=\pm2(t_0-t_2),\mu,h) \nonumber \\
  && \qquad \longrightarrow
  {\cal H}(-t_2,-t_0,U=\pm2(t_0-t_2),\mu',-h) +
	(\mu'-\mu)\ L\ .
\end{eqnarray}
Here $\mu' = 2(t_0-t_2)-\mu$ and $L$ is the number of lattice sites.

The transformation
\begin{equation}
  T_2:\quad c_{i\sigma} \rightarrow (-1)^i c_{i\sigma}\ ,
      \quad \sigma=\uparrow, \downarrow
\end{equation}
changes the sign of the single particle dispersion resulting in
\begin{equation}
  {\cal H}(t_0,t_2,U=\pm2(t_0-t_2),\mu,h) \rightarrow
  {\cal H}(-t_0,-t_2,U=\pm2(t_0-t_2),\mu,h)\ .
\end{equation}

Applying both $T_1$ and $T_2$ the sign in the first of
Eqs.~(\ref{eq:solvpt}) is reversed (see Fig.~\ref{fig:param})
\begin{eqnarray}
  &&{\cal H}(t_0,t_2,U=\pm2(t_0-t_2),\mu,h) \nonumber \\
  && \qquad \longrightarrow
  {\cal H}(t_2,t_0,U=\mp2(t_2-t_0),\mu',-h)\ .
\end{eqnarray}
Hence $t_2$ and $t_0$ are interchanged and at the same time the
electronic density is changed from $n_e$ to $2-n_e$. As will be seen later
the Bethe Ansatz solution in the region $t_0,t_2>0$ extends throughout
the interval $0\le n_e < 2$. Hence it is sufficient to consider the model
with $U=+2(t_0-t_2)$ in this region.

As mentioned above the model can be constructed in the framework of the
Quantum Inverse Scattering method based on a irreducible representation of
the algebra $gl(2|1)$. This is reflected in additional invariances  of the
Hamiltonian: apart from the $SU(2)$ spin and number operator
\begin{eqnarray}
  &&S^z={1\over 2} \sum_{i=1}^L (n_{i\downarrow}-n_{i\uparrow})\ ,
    \quad
    S^+=\sum_{i=1}^L c_{i,\uparrow}^\dagger c_{i,\downarrow}\ ,
    \quad
    S^-=\left(S^+\right)^\dagger\ , \nonumber \\
  &&N_e=\sum_{i=1}^L (n_{i\uparrow} + n_{i\downarrow})
\end{eqnarray}
which commute with the Hamiltonian for vanishing magnetic field there are
four additional supersymmetric generators \cite{bglz:95}, namely
\begin{equation}
 Q_\sigma = \sum_{i=1}^L (-1)^i c_{i,\sigma} \Bigl({t_1\over t_0}
 (1-n_{i,-\sigma})+n_{i,-\sigma} \Bigr)\ ,
 \qquad \sigma=\uparrow,\downarrow
\end{equation}
and their Hermitean conjugates $Q_\sigma^\dagger$ satisfying commutation
relations
\begin{equation}
    \left\{ Q_\uparrow,Q_\downarrow\right\} =0\ , \qquad
    Q_\sigma^2=0\ , \qquad
    \left[ {\cal H},Q_{\sigma} \right] = (\mu-2t_0+\sigma h)Q_{\sigma}\ .
 \label{eq:qf}
\end{equation}
Fixing the potentials to $\mu=2t_0$, $h=0$ one obtains the supersymmetric
model of Ref.~\cite{bglz:95} (up to the unitary transformation $T_1 T_2$).

It is important to identify the full symmetry of the model since it is well
known that the {\em Bethe-Ansatz} states are all highest weight states in
this algebra and hence not complete \cite{fata:84}, i.e.\
$S^+|\Psi_{Bethe}\rangle = 0 = Q_\sigma |\Psi_{Bethe}\rangle$. Only after
complementing the Bethe Ansatz states with those obtained by successive
application of $S^\pm$ and $Q_\sigma^\dagger$ the complete set of
eigenfunctions is found. We shall come back to this question at the
Appendix. Note, that as a consequence of (\ref{eq:qf}) the number of
particles in the states belonging to one $gl(2|1)$ multiplet range from
the number $N_{BA}$ in the {\em Bethe-Ansatz} state to $N_{BA}+2$. Hence,
in the thermodynamic limit investigated below particle densities
$0\le n_e<2$ can be studied directly.

\section{Bethe Ansatz solution in the thermodynamic limit}
Despite the derivation of the Hamiltonian (\ref{eq:Hamil}) in the framework
of the Quantum Inverse Scattering method the spectrum of the model (which
is obtainable in principle by means of the algebraic Bethe Ansatz) has not
been found in \cite{bglz:95}. The difficulty here is the complicated
representation theory for the superalgebra $gl(2|1)$.  On the other hand,
it is straightforward to determine the spectrum using the {\em coordinate}
Bethe Ansatz: for models possessing internal symmetries as the one
considered here the Schr\"odinger equation is solved with the Ansatz
\cite{yang:67}
\begin{equation}
   \Psi(X_Q) = \sum_P A_{\sigma_1,\ldots,\sigma_N}(P|Q)
   \exp\left\{ i \sum_{j=1}^{N} k_{p_j} x_j \right\}
\end{equation}
where $Q=\left\{q_1,\ldots,q_N\right\}$ and
$P=\left\{p_1,\ldots,p_N\right\}$ are permutations of the integers
$\left\{1,\ldots,N\right\}$ and $Q$ is chosen such that
$X_Q=\left\{x_{q_1}<x_{q_2}<\ldots<x_{q_N}\right\}$. The coefficients
$A(P|Q)$ from regions different than $X_Q$ are connected with each other by
elements of the two particle ${\cal S}$-matrix
\begin{equation}
   {\cal S}(k_1,k_2) = \frac{\vartheta(k_1) - \vartheta(k_2) + ic P_{12}}
	      {\vartheta(k_1) - \vartheta(k_2) + ic}
   \label{smat}
\end{equation}
($P_{12}$ is a spin permutation operator). Here the charge rapidities
$\vartheta_j$ are related to the single particle quasimomenta $k_j$ by
$\vartheta(k) = {1\over2}\tan(k/2)$ and the dependence on the system
parameters (\ref{eq:solvpt}) is incorporated in the parameter
$c=(t_0-t_2)/t_2$ (varying in the intervals $-1<c<0$ and $0<c<\infty$).
The $A(P|Q)$ are determined in a second Bethe Ansatz for an inhomogeneous
six vertex model resulting in the Bethe Ansatz equations (BAE)
\begin{eqnarray}
 \left( \frac{\vartheta_j-{i\over 2}}{\vartheta_j+{i\over 2}} \right)^L &=&
 \prod^M_{\alpha=1} \frac {\vartheta_j-\lambda_\alpha+i{c\over 2}}
 {\vartheta_j-\lambda_\alpha-i{c\over 2}}\ ,
 \quad j=1,\ldots,N_e
 \nonumber \\
 \prod^{N_e}_{j=1} \frac{\lambda_\alpha-\vartheta_j+i{c\over 2}}
 {\lambda_\alpha-\vartheta_j -i{c\over 2}} &=&
  - \prod^M_{\beta=1} \frac{\lambda_\alpha-\lambda_\beta+i c}
 {\lambda_\alpha-\lambda_\beta-ic}\ ,
 \quad \alpha=1,\ldots,M
 \label{eq:Bae}
\end{eqnarray}
The length $L$ of the system is assumed to be even and $N_e$ and $M$ are
the numbers of electrons and spin-$\downarrow$ electrons,
respectively. Given a solution of (\ref{eq:Bae}) the eigenvalue of
(\ref{eq:Hamil}) in the corresponding state is
\begin{equation}
   E=(2 t_0 - \mu) N_e - h \left({N_e\over 2}-M \right)
     -t_0 \sum_{j=1}^{N_e}{1\over \vartheta_j^2+{1\over4}}\ .
\label{eq:E-BA}
\end{equation}

Solving (\ref{eq:Bae}) one has to distinguish the two different regions
$c>0$ and $c<0$ as the character of the $\vartheta-\lambda$ solutions in
these two cases is completely different. Note that the sign of the on-site
Coulomb coupling $U$ is the same as that of $c$ because of
(\ref{eq:solvpt}). Hence the situation is very similar to the
Hubbard-model.  Below we introduce some functions and their Fourier
transforms that will be used in the following ($y>0$):
\begin{eqnarray}
 a_y(x) &=& {1\over 2\pi} \int_{-\infty}^{\infty} dk\ e^{-ikx}
            e^{-{y \over 2}|k|}
         =  {1\over2\pi}\ {y \over x^2+y^2/4} \nonumber \\
 s_y(x) &=& {1\over 2\pi} \int_{-\infty}^{\infty} dk\ e^{-ikx}
 	    {1 \over 2\cosh({yk/2})}
         =  {1\over2y\ \cosh(\pi x/y)} \\
 R_y(x) &=& (a_y*s_y)(x)
         =  {1\over 2\pi y} \hbox{Re} \left(\psi\Bigl(1+i{x\over 2y}\Bigr)
                                   -\psi\Bigl({1\over 2} +i{x\over 2y}\Bigr)
                              \right) \nonumber
\end{eqnarray}
where $(a*b)(x)=\int dz\ a(x-z)b(z)$ denotes a convolution and $\psi$ is
the Digamma-function.

\subsection{Repulsive case ($c>0$)}
In this case the solutions of (\ref{eq:Bae}) consist of real $\vartheta_j$
while the spin rapidities are known to be arranged in bound states of
uniformly spaced sets of complex $\lambda_\alpha$, so called $n$-{\em
strings}:
\begin{equation}
 \lambda_\alpha^{n,j}=\lambda_\alpha^n+i(n+1-2j){c\over 2} \quad
 j=1,2,\dots,n
\label{eq:strings}
\end{equation}
In the thermodynamic limit ($L\rightarrow \infty $ with particle density
${N_e\over L}$ and magnetization ${M\over L}$ being fixed) the solutions of
the BAE (\ref{eq:Bae}) can be described in terms of densities
$\rho(\vartheta)$ for charge rapidities and $\rho_h(\vartheta)$ for the
corresponding holes. Similarly, one introduces density distributions
$\sigma_n$ ($\sigma_{n,h}$) for the $n$-strings of spin rapidities (and
corresponding holes). Using standard procedures one obtains the following
system of coupled linear integral equations from the BAE (\ref{eq:Bae})
\begin{eqnarray}
 \rho+\rho_h &=& a_1+R_c*\rho-s_c *\sigma_{1,h} \nonumber\\
 \sigma_1+\sigma_{1,h} &=& s_c * (\sigma_{2,h}+\rho) \label{eq:densr}\\
 \sigma_n+\sigma_{n,h} &=& s_c * (\sigma_{n+1,h}+\sigma_{n-1,h})\ ,
  \quad n\ge 2 \nonumber
\end{eqnarray}
The intervals in which the densities are nonvanishing depend on the state
considered. The particle density and magnetization is related to $\rho$ and
$\sigma_n$ through
\begin{eqnarray}
    n_e = {N_e\over L} &=&
          \int_{-\infty}^\infty d\vartheta\ \rho(\vartheta)\ , \nonumber\\
    m_z = {1\over L}\left({N_e\over2}-M\right) &=&
          {n_e\over 2} - \sum_{n=1}^\infty \int_{-\infty}^\infty d\lambda\
          n\sigma_n(\lambda)\ .\nonumber
\end{eqnarray}
The energy density follows from (\ref{eq:E-BA})
\begin{equation}
    e = {E\over L} = (2t_0 -\mu) n_e -h m_z
      - 2 \pi t_0  \int_{-\infty}^\infty d\vartheta\ a_1(\vartheta)
       \rho(\vartheta)\ .
\end{equation}

The equilibrium distribution functions $\rho$ and $\sigma_n$ have to be
determined by minimization of the free-energy functional, ${\cal F}=E-TS$,
with the combinatorical entropy $S$ of a particle and hole densities
$\delta(\lambda)$ and $\delta_h(\lambda)$ given by \cite{yaya:69}
\begin{equation}
   {S_\delta\over L} = \int_{-\infty}^\infty d\lambda\
   \left\{ (\delta+\delta_h) \ln(\delta+\delta_h)
           -\delta \ln(\delta) -\delta_h \ln(\delta_h)\right\}\ .
\end{equation}
Introducing the following functions
\begin{equation}
   \varepsilon_c(\vartheta)=T \ln \Bigl({\rho_h \over \rho} \Bigr)\ ,
   \quad
   \varepsilon_n(\lambda)=T \ln \Bigl({\sigma_{n,h} \over \sigma_n}\Bigr)\
\label{eq:dresse}
\end{equation}
and considering $\rho$ and $\sigma_{n,h}$ as independent functions we
obtain by variation of ${\cal F}$ the following nonlinear integral
equations for the functions $\varepsilon_\alpha$
\begin{eqnarray}
    \varepsilon_c &=& 2t_0-\mu-2\pi t_0 a_1
                   + Ts_c * \ln(n(\varepsilon_1))
                   + TR_c * \ln(n(-\varepsilon_c))
	\nonumber \\
    \varepsilon_1 &=& -Ts_c * \ln(n(\varepsilon_2))
                      +Ts_c * \ln(n(-\varepsilon_c))
	\label{eq:rr}\\
    \varepsilon_n &=& -Ts_c * \left( \ln(n(\varepsilon_{n+1}))
                                    +\ln(n(\varepsilon_{n-1}))\right)\ ,
    \quad n \ge 2
	\nonumber
\end{eqnarray}
with the distribution function
\begin{equation}
   n(\varepsilon)=\left({1+e^{\varepsilon\over T}}\right)^{-1}\ .
\label{eq:n}
\end{equation}
Eqs.~(\ref{eq:rr}) have to be solved with the asymptotic boundary
condition:
\begin{equation}
   \lim_{n \to \infty} {\varepsilon_n \over n} = h
\label{eq:lh}
\end{equation}
The free energy density is given by
\begin{equation}
 f=T \int_{-\infty}^\infty d\vartheta\ a_1(\vartheta) \ln(n(-\varepsilon_c
 (\vartheta)))\ .
\end{equation}
This shows that the functions $\varepsilon_\alpha$ are to be identified as
renormalized (``dressed'') energies of the single particle excitations in
the system.
\subsection{Attractive case ($-1<c<0$)}
In this regime one has---in addition to the real charge rapidities and
strings of spin rapidities considered in the repulsive case---pairs of
complex conjugated $\vartheta_j^\pm$ coupled to a real $\lambda_j$ as
solutions of the BAE (\ref{eq:Bae})
\begin{equation}
   \vartheta_j^\pm = \lambda'_j \pm {i|c| \over 2}
\label{eq:tlpairs}
\end{equation}
Note that for $c\to-1$ (\ref{eq:Bae}) would coincide with the BAE of the
$t$--$J$-model obtained in \cite{schl:87}, however this value is out of
the range accessible for this model.  Following the same programm as in
the repulsive case we obtain in the thermodynamic limit
\begin{eqnarray}
    \rho+\rho_h &=& s_{|c|} *\sigma'_{h} + s_{|c|}*\sigma_{1,h}
	\nonumber\\
    \sigma'+\sigma'_h &=& a_{1-|c|}+R_{|c|}*\sigma'_h-s_{|c|}*\rho
	\nonumber \\
    \sigma_1+\sigma_{1,h} &=& s_{|c|} * (\sigma_{2,h}+\rho)
	\nonumber\\
    \sigma_n+\sigma_{n,h} &=& s_{|c|} * (\sigma_{n+1,h}+\sigma_{n-1,h})\quad
     n\ge 2
\label{eq:sigs}
\end{eqnarray}
where $\sigma'(\lambda)$ and $\sigma'_h(\lambda)$ are the distribution
function for the paired rapidities (\ref{eq:tlpairs}) and corresponding
holes.

Particle density and magnetization of the state corresponding to a solution
of (\ref{eq:sigs}) are given by the follwoing expressions:
\begin{eqnarray}
    n_e &=& \int_{-\infty}^\infty d\vartheta\ \rho(\vartheta)
    +2 \int_{-\infty}^\infty d\lambda\ \sigma'(\lambda)\ ,
	\nonumber\\
    m_z &=& {1\over 2} \int_{-\infty}^\infty d\vartheta\ \rho(\vartheta)
         - \sum_{n=1}^\infty \int_{-\infty}^\infty d\lambda\
		n \sigma_n (\lambda)\ ,
\nonumber
\end{eqnarray}
and the energy density is
\begin{eqnarray}
 e &=& (2t_0-\mu) n_e -h m_z
     - 2 \pi t_0  \int_{-\infty}^\infty d\vartheta\ a_1(\vartheta)
	 \rho(\vartheta) \nonumber \\
 && -2 \pi t_0 \int_{-\infty}^\infty d\lambda\
	\left(a_{1+|c|}+a_{1-|c|}\right)(\lambda)
	\sigma'(\lambda)\ .
\end{eqnarray}
Minimizing the free energy (with $\sigma'_h$ as additional independent
function) and defining the dressed energy of the paired rapidities as
\begin{equation}
	\varepsilon_p(\lambda)=T \ln \Bigl({\sigma'_h \over \sigma} \Bigr)
\end{equation}
we obtain the thermodynamic Bethe Ansatz equations for the attractive case
\begin{eqnarray}
   \varepsilon_c &=& Ts_{|c|} * \ln(n(\varepsilon_1))-Ts_{|c|} *
    \ln(n(\varepsilon_p))
	\nonumber\\
   \varepsilon_p &=& 4t_0-2\mu-2\pi t_0 (a_{1+|c|}+a_{1-|c|})
	- T \left(a_{2|c|}*\ln(n(-\varepsilon_p))
                 +a_{|c|}*\ln(n(-\varepsilon_c))
            \right)
	\nonumber\\
   \varepsilon_1 &=& -Ts_{|c|} * \left( \ln(n(\varepsilon_2))
					+ \ln(n(-\varepsilon_c))\right)
	\nonumber \\
   \varepsilon_n &=& -Ts_{|c|} * \left(\ln(n(\varepsilon_{n+1}))
				    +\ln(n(\varepsilon_{n-1}))\right)\ ,
	 \quad n \ge 2
\label{eq:zz}
\end{eqnarray}
to be solved with the field boundary condition (\ref{eq:lh}).  The free
energy density is given by
\begin{equation}
 f=T \int_{-\infty}^\infty d\vartheta\
   \left(a_1(\vartheta)\ln (n(-\varepsilon_c))
      +(a_{1-|c|}+a_{1+|c|})\ln (n(-\varepsilon_p))
   \right)\ .
\end{equation}

\section{Ground state and excitations at $T=0$}
Now we want to examine properties of the zero temperature ground state for
the two cases. In this limit the distribution function $n$ (\ref{eq:n}) in
the thermodynamic BAE reduces to
\begin{equation}
	\lim_{T \to 0} T\ln(n(\delta)) = - \delta^+
\label{eq:lt}
\end{equation}
where $\delta^+>0$ and $\delta^- <0$ are the positive and negative parts of
the function $\delta=\delta^+ +\delta^-$, respectively. At the same time it
is clear that the ground state configuration corresponds to the filling of
all states with negative dressed energy $\varepsilon_\alpha$.
\subsection{Repulsive case ($c>0$)}
{}From (\ref{eq:rr}) we find that $\varepsilon_{n>1}(\lambda)>0$ for all
$\lambda$. Using the asymptotic condition (\ref{eq:lh}) we obtain from
(\ref{eq:rr}) with (\ref{eq:lt})
\begin{equation}
   \left( \begin{array}{c} \varepsilon_c \\
			   \varepsilon_1 \end{array} \right) =
   \left( \begin{array}{c} 2t_0-\mu-2 \pi t_0 a_1-{h\over 2} \\
			   h \end{array} \right)+
   \left( \begin{array}{cc} 0 & a_c \\ a_c & -a_{2c} \end{array} \right)*
   \left( \begin{array}{c} \varepsilon_c^- \\
			   \varepsilon_1^- \end{array} \right)
 \label{eq:de1}
\end{equation}
As in \cite{taka:72} one can prove that $\varepsilon_c(\vartheta)$ and
$\varepsilon_1(\lambda)$ are monotonically increasing functions of
$|\vartheta|$ and $|\lambda|$. Consequently, they are negative in the
intervals $[-Q,Q]$ and $[-B,B]$. For $h=0$ two possible ground state
configurations are to be considered: the ferromagnetic state ($M=0$ for
$n_e\le1$ and $M=N_e-L$ for $n_e>1$) and the antiferromagnetic one
($M=N_e/2$). The energy of the ferromagnetic state at fixed density $n_e$
is simply
\begin{equation}
   e_{FM} = -{2t_0} \left\{
     \begin{array}{ll}
	 {1\over\pi}\sin\pi n_e & \hbox{for~} n_e\le1 \\[8pt]
	 {1\over 1+c}\left( {1\over\pi}\sin\pi n_e - c(n_e-1)\right) &
				 \hbox{for~} n_e>1
     \end{array}\right.
    \label{eq:E0-FM}
\end{equation}
For the antiferromagnetic state one obtains from (\ref{eq:densr}) that it
corresponds to a filled band of 1-strings, i.e.\ $B=\infty$. Hence
$\sigma_1$ can be eliminated by Fourier transform and (\ref{eq:densr})
simplifies to:
\begin{equation}
	\rho(\vartheta)= a_1(\vartheta)
		+\int_{-Q}^Q d\vartheta'\
			R_c(\vartheta-\vartheta') \rho(\vartheta')\ .
\label{eq:rog}
\end{equation}
Varying $Q$ one obtains any filling between $n_e=0$ and $n_e=2$: For small
$Q$ (\ref{eq:rog}) can be solved by iteration and for $Q\to \infty$ using
Wiener Hopf techniques \cite{yaya:66} with the result
\begin{equation}
	n_e = \left\{
  \begin{array}{ll}
	{4Q\over \pi} + 8 {\ln2 \over \pi^2 c}Q^2+O(Q^3)\ ,
               & \hbox{for~} Q\to 0\  \\[8pt]
	2-{2(c+1)\over \pi Q} \left(1+c{\ln(Q) \over 2\pi Q}\right)
	+O\left({1\over Q^2}\right)\  &
		 \hbox{for~} Q\to \infty\ .
  \end{array}
  \right.
\end{equation}
In the low density limit we find that the ground state of the system is
indeed antiferromagnetic. The energy difference to (\ref{eq:E0-FM}) is
\begin{equation}
	 e_{FM} - e_0= \left\{
  \begin{array}{ll}
	{\pi^2\ln2 \over3c}\  n_e^4 + O(n_e^5)\ ,
               & \hbox{for~} n_e \to 0\  \\[8pt]
	{4t_0 \over 1+c} (2-n_e)+o\left((2-n_e)^2\right)\  &
		 \hbox{for~} n_e \to 2\ .
  \end{array}
  \right.
  \label{eq:EAFM}
\end{equation}
In Figure~\ref{fig:gsrep} we present numerical data for the dependence of
the (antiferromagnetic) ground state energy of the system as compared to
the ferromagnetic one for various values of the parameter $c$.\footnote{%
In Ref.~\cite{karn:94} the ground state is claimed to be {\em ferro\/}magnetic
in this regime for densities $n_e<1$. As is clear from Eq.~(\ref{eq:EAFM}),
this is not correct.}

In the free fermion limit $c \to 0$ (\ref{eq:rog}) simplifies to
\begin{equation}
	\rho(\vartheta)= a_1(\vartheta)
	+{1\over 2}\int_{-Q}^Q d\vartheta'\
	\delta(\vartheta-\vartheta') \rho(\vartheta')
\end{equation}
and the ground state energy is the expected result for this system
\begin{equation}
	e_0=-{4t_0\over \pi} \sin({\pi n_e \over 2})\ .
\label{eq:egg1}
\end{equation}
In the strong coupling limit $c \to\infty$ corresponding to $t_1=t_2=0$,
$t_3=-t_0$, $U=2t_0$, the groundstate is degenerate with the ferromagnetic
state (\ref{eq:E0-FM}).

There are two types of excitations that are to be considered in the low
energy sector: first, there are objects carrying charge (`holons')
corresponding to particle or hole like excitations in the ground state
configuration of charge rapidities with energy $|\varepsilon(\vartheta)|$.
Furthermore, there are spin carrying objects (`spinons') corresponding to
holes in the distribution of real spin rapidities. From (\ref{eq:de1})
their energy is found to be
\begin{equation}
   \varepsilon_1(\lambda) = {1\over2c}\int_{-Q}^Q d\vartheta\
    {|\varepsilon_c(\vartheta)|\over
     \cosh{\pi\over c}(\lambda-\vartheta)}\ .
\end{equation}
The physical excitations (for even particle number $N_e$) are even numbers
of these objects forming a continuum of spin waves without gap. The
energies of the spin rapidity strings of length $n>1$ vanish.

Increasing the magnetic field the magnetization grows until it reaches its
saturation value $1\over 2$ at the critical field $h_c$. For $h=h_c$ the
interval for the $\lambda$-integration vanishes, i.e.\ $B=0$ (corresponding
to $\varepsilon_1(\lambda=0)=0$). For $0\le n_e \le 1$ we find
\begin{equation}
 h_c={8t_0\over \pi} \frac{c(4Q^2+1)\arctan(2Q)-(4Q^2+c^2)\arctan({2Q\over c})}
     {(c^2-1)(4Q^2+1)}
\end{equation}
with $Q={1\over 2}\tan(\frac{\pi n_e} {2})$. In the limiting cases
considered above this expression becomes
\begin{equation}
	\lim_{Q \to 0}h_c=0\ ,\qquad \lim_{Q \to \infty}h_c=4t_2\ ,
	\qquad \lim_{c \to 0}h_c=4t_0 \sin ^2({\pi n_e \over 2}) .
\end{equation}

\subsection{Attractive case ($-1<c<0$)}
Due to (\ref{eq:zz}) the dressed energies of the spin rapidities
$\varepsilon_{n\ge 1}(\lambda)$ are always positive. Performing the
limit $T\to 0$ in (\ref{eq:zz}) with Eqs.~(\ref{eq:lt}) and (\ref{eq:lh})
we obtain
\begin{equation}
    \left( \begin{array}{c} \varepsilon_p \\
			    \varepsilon_c \end{array} \right) =
\left( \begin{array}{c} 4t_0-2\mu-2 \pi t_0 (a_{1+|c|}+a_{1-|c|})\\
 2t_0-\mu-2\pi t_0 a_1 -{h\over 2} \end{array} \right)-
    \left( \begin{array}{cc} a_{2|c|} & a_{|c|} \\
                             a_{|c|} & 0 \end{array} \right) *
    \left( \begin{array}{c} \varepsilon_p^- \\
                            \varepsilon_c^- \end{array} \right)\ .
 \label{eq:de2}
\end{equation}
As in the repulsive regime one can prove that $\varepsilon_c(\vartheta)$
and $\varepsilon_p(\lambda)$ are monotonically increasing functions of the
modulus of their arguments. Hence they are negative in the regions $[-Q,Q]$
and $[-B,B]$. Again we find that the ground state of the system is
antiferromagnetic for $h=0$ (see Fig.~\ref{fig:gsatt}). In this regime the
ground state configuration consists of paired rapidities only ($Q=0$).
Their density is obtained from (\ref{eq:sigs}) which simplifies to
\begin{equation}
    \sigma'(\lambda) = a_{1-|c|}(\lambda) + a_{1+|c|}(\lambda)
		-\int_{-B}^B d\mu\ a_{2|c|}(\lambda-\mu)\sigma'(\mu)
\label{eq:sig}
\end{equation}
Again, this is the ground state configuration for any filling $0\le
n_e\le2$ since
\begin{equation}
 n_e= \left\{
  \begin{array}{ll}
    {8B\over \pi}\ {2\over 1-c^2}+O(B^2) &
	\hbox{for~} B\to 0\ , \\[8pt]
    2-{2(1-|c|)\over \pi B} (1-{|c|\ln(B) \over 2\pi B} )+O({1\over B^2} ) &
	\hbox{for~} B\to \infty\ .
  \end{array}
   \right.
\end{equation}

For $c \to 0$ (\ref{eq:sig}) simplifies to:
\begin{equation}
 \sigma'(\lambda)= 2a_1(\lambda)-\int_{-B}^B  d\mu\ \delta(\lambda-\mu)
		\sigma'(\mu)
\end{equation}
and we obtain for the ground state energy (\ref{eq:egg1}) of the free
fermion system.

{}From Eq.~(\ref{eq:de2}) we obtain the dressed energy for excitations
corresponding to real charge rapidities:
\begin{equation}
 \varepsilon_c(\vartheta) =-{h\over 2}
      + {1\over |c|} \int_B^\infty d\lambda\
	{1\over\cosh((\vartheta-\lambda)\pi /c)}\
	 \varepsilon_p(\lambda)
\label{eq:gg}
\end{equation}
Note that for vanishing magnetic field there is a gap $\Delta_c
(n_e)=\varepsilon_c (0)$ for the creation of unpaired electrons. $\Delta_c$
is a monotonically falling function of $n_e$ with its maximum at $\Delta_c
(0)={4t_0 c^2 \over 1-c^2}>0$ (see Figure~\ref{fig:gap}). The only massless
excitations in this regime are charge density waves corresponding to
excitations within the band $\varepsilon_p(\lambda)$.

In an external magnetic field the nature of the excitations in the system
change: For $h_{c1}=2\Delta_c$ the gap for charge excitations closes and
for $h_{c2}$ the system undergoes a transition into the saturated
ferromagnetic ground state.  The latter corresponds to $B=0$ (or,
equivalently $\varepsilon_p(0)=0$).  For $0 \le n_e\le 1$ we find:
\begin{equation}
 h_{c2}={8t_0 c^2\over 1-c^2}+{8t_0\over \pi} \frac{(4B^2+c^2)
  \arctan({2B\over |c|})-|c|(4B^2+1)\arctan(2B)} {(c^2-1)(4B^2+1)}
\end{equation}
with $B={1\over 2}\tan(\frac{\pi n_e} {2})$. We obtain the following limiting
cases:
\begin{equation}
 \lim_{B \to \infty}h_{c2}=4t_2 \qquad \lim_{B \to 0}h_{c_2}=2\Delta_c (0),
  \qquad \lim_{c\to0}{h_{c_2}}=4t_0\sin ^2({\pi n_e \over 2}).
\end{equation}
The nonvanishing of $h_{c2}$ in the low density limit is a direct
consequence of the gap of $\varepsilon_c$.

\section{Finite size corrections and critical exponents}
We now want to study the finite size corrections of the spectrum to discuss
the asymptotic behavior of correlation functions. Again the repulsive and
the attractive case are completely different and have to be treated separately.
\subsection{Repulsive case ($c>0$)}
As found in Section~4.1 above for $T=0$ the ground state and low lying
excitations are obtained from solutions of the BAE (\ref{eq:Bae}) with real
$\vartheta$'s and $\lambda$'s.  Hence, we have the same situation as in the
repulsive Hubbard model and following the procedure in Ref.~\cite{woyn:89}
we obtain the finite size corrections of the ground state energy as
\begin{equation}
E_0-Le_0=-{\pi\over 6L}(v_c+v_s)+o\left({1\over L}\right)\ ,
\end{equation}
where $v_c$ and $v_s$ are the Fermi velocities of charge and spin density
waves, respectively:
\begin{equation}
 v_c={1\over 2\pi \rho(Q)} \varepsilon'_c(Q),\quad
 v_s={1\over 2\pi \sigma_1 (B)} \varepsilon'_1 (B)\ .
\end{equation}
Similarly the energies and momenta of the low lying excitations are given
by
\begin{eqnarray}
   E({\bf \Delta N,D})-Le_0&=&\frac{2\pi}{L}\left[
	v_c(\Delta_c^+ +\Delta_c^- )+v_s (\Delta_s^+ +\Delta_s^-)\right]
	+o\left({1\over L}\right) \nonumber \\
   P({\bf \Delta N,D})-P_0&=&\frac{2\pi}{L}\left[
	\Delta_c^+ - \Delta_c^- +\Delta_s^+  - \Delta_s^-\right]
     +2D_c{\cal P}_{F,\uparrow}+2(D_c+D_s){\cal P}_{F,\downarrow}
\label{eq:eex}
\end{eqnarray}
with the conformal dimensions
\begin{eqnarray}
  2\Delta_c^\pm ({\bf \Delta N,D})&=&
	\left(Z_{cc}D_c+Z_{sc}D_s \pm \frac {Z_{ss}\Delta
	N_c-Z_{cs}\Delta N_s} {2 \det(Z)}\right)^2+2N_c^\pm \nonumber \\
  2\Delta_s^\pm ({\bf \Delta N,D})&=&
	\left(Z_{cs}D_c+Z_{ss}D_s \pm \frac {Z_{cc}\Delta
	N_s-Z_{sc}\Delta N_c} {2 \det(Z)}\right)^2+2N_s^\pm \ ,
\label{eq:codi1}
\end{eqnarray}
and the Fermi momenta ${\cal P}_{F,\uparrow (\downarrow)}={\pi\over 2}
(n_e\pm 2m_z)$ of spin up (down) electrons. The elements of the two
component vectors ${\bf \Delta N}$ and {\bf D} characterize the excited
state: ${\bf \Delta N}$ has integer components denoting the change of the
number of electrons and down spins with respect to the ground state.~$D_c$
and $D_s$ describe the deviations from the symmetric ground state
distributions.  They are integers or half-odd integers depending on the
parities of $\Delta N_c$ and $\Delta N_s$:
\begin{equation}
 D_c={\Delta N_c+\Delta N_s \over 2} \quad \mbox{mod} \quad 1\ ,\quad
 D_s={\Delta N_c\over 2}
 \quad \mbox{mod} \quad 1
 \label{eq:dcs}
\end{equation}
The matrix
\begin{equation}
Z= \left( \begin{array}{cc} Z_{cc} & Z_{cs} \\
  Z_{sc} & Z_{ss} \end{array} \right)=
  \left( \begin{array}{cc} \xi_{cc}(Q) & \xi_{sc} (Q) \\
  \xi_{cs} (B) & \xi_{ss} (B) \end{array} \right)^\top
\label{eq:z}
\end{equation}
parametrizing the conformal dimensions (\ref{eq:codi1}) is given in terms
of the so called dressed charge matrix which satisfies a linear integral
equation similar to (\ref{eq:de1})  for the dressed energies
\begin{equation}
\left( \begin{array}{cc} \xi_{cc}(\vartheta) & \xi_{sc}(\vartheta) \\
  \xi_{cs}(\lambda) & \xi_{ss}(\lambda) \end{array} \right)=
\left( \begin{array}{cc} 1 & 0 \\ 0 & 1 \end{array} \right)+
 \left( \begin{array}{cc} 0 & a_c \\ a_c & -a_{2c} \end{array} \right)*
 \left( \begin{array}{cc} \xi_{cc}(\vartheta) & \xi_{sc}(\vartheta) \\
  \xi_{cs}(\lambda) & \xi_{ss}(\lambda) \end{array} \right)\ .
\end{equation}

As shown above the ground state at vanishing magnetic field corresponds to
$B=\infty$ and with the aid of the Wiener--Hopf method \cite{woyn:89}
(\ref{eq:z}) simplifies to
\begin{equation}
Z= \left( \begin{array}{cc} Z_{cc} & Z_{cs} \\
  Z_{sc} & Z_{ss} \end{array} \right)=
  \left( \begin{array}{cc} \xi_c(Q) & 0 \\
  {1\over 2}\xi_c (Q) & {\sqrt{2} \over 2} \end{array} \right) \label{eq:zhn}
\end{equation}
where $\xi_c$ is defined as the solution of the following scalar integral
equation
\begin{equation}
\xi_c(\vartheta)=1+\int_{-Q}^Q d\vartheta' R_c(\vartheta-\vartheta')
  \xi_c(\vartheta').
\label{eq:xi}
\end{equation}
As for the density one can solve (\ref{eq:xi}) near $Q=0$ and $Q=\infty$
with the result
\begin{equation}
   \xi_c(Q)=\left\{
 \begin{array}{ll}
       1+{2 \ln2 \over \pi c}Q+O(Q^2), & \hbox{for~} Q\to 0\ \\[8pt]
       \sqrt{2}(1-{c\over 4 \pi Q}) +o\left({1\over Q}\right)\ &
       \hbox{for~} Q\to \infty\ .
 \end{array}
 \right.
\end{equation}
Hence the range of variation for the exponents determining the long
distance asymptotics of the equal time correlators is the same as in the
Hubbard- and $t$--$J$- model \cite{frko:90,kaya:91}. Introducing
$\theta=2\xi_c^2 (Q)$ the singularity of the momentum distribution function
at the Fermi point is found to be
\begin{eqnarray}
   n_\sigma(k) &\sim& \int dx\ e^{-ikx}
	\langle c_{x,\sigma}(t=0^+) c_{0,\sigma}(t=0) \rangle \nonumber \\
   &\propto& \hbox{sgn}(k-{\cal P}_F) |k-{\cal P}_F|^\nu,\qquad
	\nu = {1\over\theta} + {\theta\over16} - {1\over2}
\end{eqnarray}
with a variation of the exponent $\nu$ in the interval $0\le\nu\le{1\over
8}$ which shows the expected Luttinger liquid behaviour of this system.

Similarly, we obtain for the density density and singlet pair correlation
functions (${\cal P}_{F,\uparrow} = {\cal P}_{F,\downarrow}
\equiv {\cal P}_{F}$):
\begin{eqnarray}
 G_{nn}(x) &=& \langle (n_{x\uparrow}+n_{x\downarrow}) (n_{0\uparrow}+
 n_{0\downarrow}) \rangle \nonumber \\
 &\sim& n_e^2+A_1 \cos(2{\cal P}_F x+\varphi_1) x^{-(1+\theta /4)}
  +A_2 \cos(4{\cal P}_F x+\varphi_2) x^{-\theta}+A_3 x^{-2} \\
 G_{p}^{(0)}(x) &=& \langle c_{x+1,\uparrow}^\dagger c_{x,\downarrow}^\dagger
 c_{1,\downarrow} c_{0,\uparrow}\rangle \sim
	A \cos(2{\cal P}_F x+\varphi)x^{-(4/\theta+\theta/4)} \nonumber
\end{eqnarray}
The leading order of the density--density correlator is given by the $A_1$
term with $3/2 < 1+\theta/4 < 2$. Comparing this with the leading term
of the singlet--pair correlator $5/2 > 4/\theta+\theta/4 >2$ we see that
density fluctuations are dominant.

In Figure~\ref{fig:dcrep} we show lines of constant $\xi_c(Q)$ (hence
identical critical behavior) in the $n_e$--$c$ parameter plane. Note that
the strong coupling result $\xi_c(Q)=1$ is found for less than half filling
only. Beyond half filling the density dependence of the dressed charge is
for $c =\infty$
\begin{equation}
   \xi_c=\left\{
 \begin{array}{ll}
	n_e & \hbox{for~} n_e \to 1 \\[8pt]
	\sqrt{2}\left(1-{1\over 8}(2-n_e)\right) &
	\hbox{for~} n_e \to 2 \ .
 \end{array}
 \right.
\end{equation}

As in \cite{frko:91} for the Hubbard model this analysis of the critical
behaviour can be extended to the case of magnetic fields. For small fields
$h<h_c$ one has to expect logarithmic singularities in the exponents while
for fields $h>h_c$ the ground state is a saturated ferromagnetic one and
spin density waves become massive giving a scalar dressed charge instead of
(\ref{eq:zhn}).

\subsection{Attractive case $c>0$}

As discussed in Section~4.2 for $T=0$ and $h<h_{c1}$ there is only one
branch of massless excitations within the band $\varepsilon_p$.\footnote{%
In the analysis of the asymptotics of correlation functions for the model
with $c=-1/2$ in \cite{karn:95} the existence of a second branch of
massless excitations in the band of real charge rapidities $\varepsilon_c$
is assumed.  However, as shown in Sect.~4.1 these have a gap for
$h<h_{c1}$. Hence the results in \cite{karn:95} are incorrect.}
The finite size corrections to the energies of the low lying excitations
are given by
\begin{eqnarray}
  E(\Delta N_p,D_p)-Le_0&=&\frac{2\pi}{L}v_p(\Delta_p^+ +\Delta_p^- )
  +o\left({1\over L}\right) \nonumber \\
  P(\Delta N_p,D_p)-P_0&=&\frac{2\pi}{L}(\Delta_p^+ - \Delta_p^- )
  +2D_p{\cal P}_F
\end{eqnarray}
with
\begin{equation}
   2\Delta_p^\pm ({\Delta N_p,D_p}) = \left(\xi_p (B) D_p \pm
   \frac {\Delta N_p}  {2 \xi_p (B)}\right)^2+2N_p^\pm
\end{equation}
and charge density wave velocity $v_p=\varepsilon'_p(Q)/(2\pi \sigma'(Q))$
The dressed charge  $\xi_p$ is  given by
\begin{equation}
\xi_p(\lambda)=1-\int_{-B}^B d\lambda' a_{2|c|}(\lambda-\lambda')
  \xi_p(\lambda'). \label{eq:xip}
\end{equation}
With the same techniques as above we obtain
\begin{equation}
   \xi_p(B)=\left\{
 \begin{array}{ll}
       1-{2 \over \pi |c|}B+O(B^2), & \hbox{for~} B\to 0\ \\[8pt]
       {\sqrt{2} \over 2}\left(1+{|c|\over 4 \pi B}\right)
        +o\left({1\over B}\right)\ &
       \hbox{for~} B\to \infty\ .
 \end{array}
 \right.
\end{equation}
The leading terms in the asymptotics of the equal time correlators as a
function of $\theta=2\xi_p^2 (B)$ are the same as in the (attractive)
Hubbard model \cite{boko:90}
\begin{equation}
 G_{nn}(x) \sim  n_e^2
	+A_1 \frac{\cos(2{\cal P}_Fx)}{x^\theta}
	+\frac{A_2}{x^2}\ , \qquad
 G_{p}^{(0)}(x) \sim x^{-1 / \theta}
\end{equation}
Comparing the leading exponents of these two correlators we see that the
correlation of pairs $(1/2 \le 1/ \theta \le 1)$ overwhelms the
density--density correlator $(2 \ge \theta \ge 1)$ fo arbitray $n_e$.
So as in the attractive Hubbard model \cite{boko:90} we can conclude that the
particles are confined in pairs which is reflected in the structure of the
{\em Bethe Ansatz} ground state configuration.

In Figure~\ref{fig:dcatt} we show lines in the $n_e$--$|c|$ parameter plane
with identical critical behavior.

For $h \ge h_{c1}$ charge and spin excitations are massless and the dressed
charge is a $2\times2$ matrix as in the repulsive case. The same situation
occurs in the attractive Hubbard model \cite{boko:90}.

\section*{Acknowledgements}
This work has been supported in part by the Deutsche
Forschungsgemeinschaft under Grant No.\ Fr~737/2--1.

\appendix

\section{Completeness of the {\em Bethe Ansatz} states}

As mentioned above the {\em Bethe Ansatz} states do not form the {\em
complete} set of eigenstates of the systen (\ref{eq:Hamil}) but are the
highest weight states of the $gl(2|1)$ superalgebra. Complementing the {\em
Bethe Ansatz} states with those obtained by the action of the $gl(2|1)$
shift operators one obtains additional eigenstates. The completeness of
this {\em extended} Bethe Ansatz has been proven (based on a string
hypothesis (\ref{eq:strings}) for the solutions of the BAE) for some models
such as the the spin $1\over2$ Heisenberg chain, the supersymmetric
$t$--$J$ model and the Hubbard model \cite{fata:84,foka:93,eks:92}.  In
this appendix we present the study of the completeness for the two-site
system together with some remarks on $L>4$.

New eigenstates of the system are generated from the {\em Bethe Ansatz}
states by acting with the total spin operators $S^-$, $S^+$ and the
supersymmetry generators $Q_\sigma$, $Q_\sigma^\dagger$. As a consequence
of the anticommutativity of the latter (\ref{eq:qf}) the resulting
multiplet contains states in the $N_e$--, $(N_e+1)$-- and $(N_e+2)$--particle
sectors which are (we suppress the spin-multiplicity):
\begin{equation}
 |\Psi_{Bethe} \rangle \stackrel{Q_{\uparrow \downarrow}^\dagger}
 {\longrightarrow}
 \left\{
 \begin{array}{c} |\Psi_{Q1} \rangle \\ |\Psi_{Q2} \rangle \end{array}
 \right\}
 \stackrel{Q_{\downarrow \uparrow}^\dagger}{\longrightarrow}
 |\Psi_{Q3} \rangle\ .
\label{app:rais}
\end{equation}
As shown above, the ground state of the model for {\em fixed} number of
particles is always a spin singlet. As a consequence of (\ref{app:rais})
it is member of a $gl(2|1)$ quartet, the same situation as in the related
supersymmetric $t$--$J$ model \cite{foka:93}.

Solving the BAE (\ref{eq:Bae}) in the simplest case of the $L=2$ system we
obtain three {\em regular} {\em Bethe Ansatz} states $|\psi_i\rangle$ with
energy $E_i$ (at the supersymmetric point $\mu=2t_0$, $h=0$):
\begin{eqnarray}
  |\psi_1 \rangle &=& |N_e=0,M=0 \rangle \equiv |0 \rangle\ ,\quad
	E_1 = 0\nonumber \\
  |\psi_2 \rangle &=& |N_e=1,M=0 \rangle \equiv |k=0,\uparrow \rangle\ ,\quad
	E_2 = -4t_0\\
  |\psi_3 \rangle &=& |N_e=2,M=1 \rangle
	\propto |\psi_{\uparrow \downarrow} \rangle
	 -|\psi_{\downarrow \uparrow} \rangle
	 +{t_1\over t_0} (|\psi_{20} \rangle + |\psi_{02}
  \rangle )\ , \quad E_3 = -4(t_0+t_2)
	\nonumber
\end{eqnarray}
with
\begin{equation}
 |\psi_{\sigma_1 \sigma_2} \rangle
     = c_{1,\sigma_1}^\dagger c_{2,\sigma_2}^\dagger |0 \rangle\ , \quad
 |\psi_{20} \rangle = c_{1,\uparrow}^\dagger  c_{1,\downarrow}^\dagger
	|0 \rangle\ ,\quad
 |\psi_{02} \rangle = c_{2,\uparrow}^\dagger  c_{2,\downarrow}^\dagger
	|0 \rangle\ .
\end{equation}
{\em Regular} Bethe Ansatz states are those corresonding to solutions of
(\ref{eq:Bae}) with {\em finite} $\vartheta$ and $\lambda$
\cite{fata:84,eks:92}.

The one- and two particle descendants of $|\psi_1 \rangle$ are found to be
the momentum $\pi$ spin-doublet $|k=\pi,\sigma \rangle$ and the spin
singlet
\begin{equation}
  |\psi_{\uparrow \downarrow} \rangle-|\psi_{\downarrow \uparrow} \rangle
  -{t_0\over t_1} (|\psi_{20} \rangle + |\psi_{02} \rangle )\ .
\end{equation}
Analogously we find the descendants of $|\psi_2 \rangle$ to be the
following (degenerate) triplet and singlet states in the two particle
sector
\begin{equation}
	|\psi_{\uparrow \uparrow}\rangle,\quad
	|\psi_{02} \rangle -  |\psi_{20} \rangle
\end{equation}
and the doublet of zero-momentum single hole states $|k_h=0,\sigma\rangle$.
Finally, $|\psi_3 \rangle$ leads to doublet of momentum $\pi$ hole states
$|k_h=\pi,\sigma \rangle$ and the completely filled state
$|\psi_{22}\rangle$.

Hence the {\em Bethe Ansatz} extended by means of the supersymmetry does
indeed give the complete spectrum of states on the two site lattice. Note,
that $|\psi_3\rangle$ is {\em always} the ground state of the two particle
sector for the range of parameters considered here. The difference between
the repulsive and attractive regime is the larger amplitude of the states
containing local pairs in the latter.

For general $L$ regular {\em Bethe Ansatz} states will exist for particle
numbers up to $2(L-1)$. Considering $L=4$ as an example one has to find 35
regular solutions of the BAE to generate a complete set of eigenstates.
Four of these are states with $N_e > L$.

\newpage

\setlength{\baselineskip}{13pt}

\newpage

\centerline{\large \bf Figures}
\begin{figure}[h]
\begin{center}
	\leavevmode
	\epsfxsize=1.00\textwidth
	\epsfbox{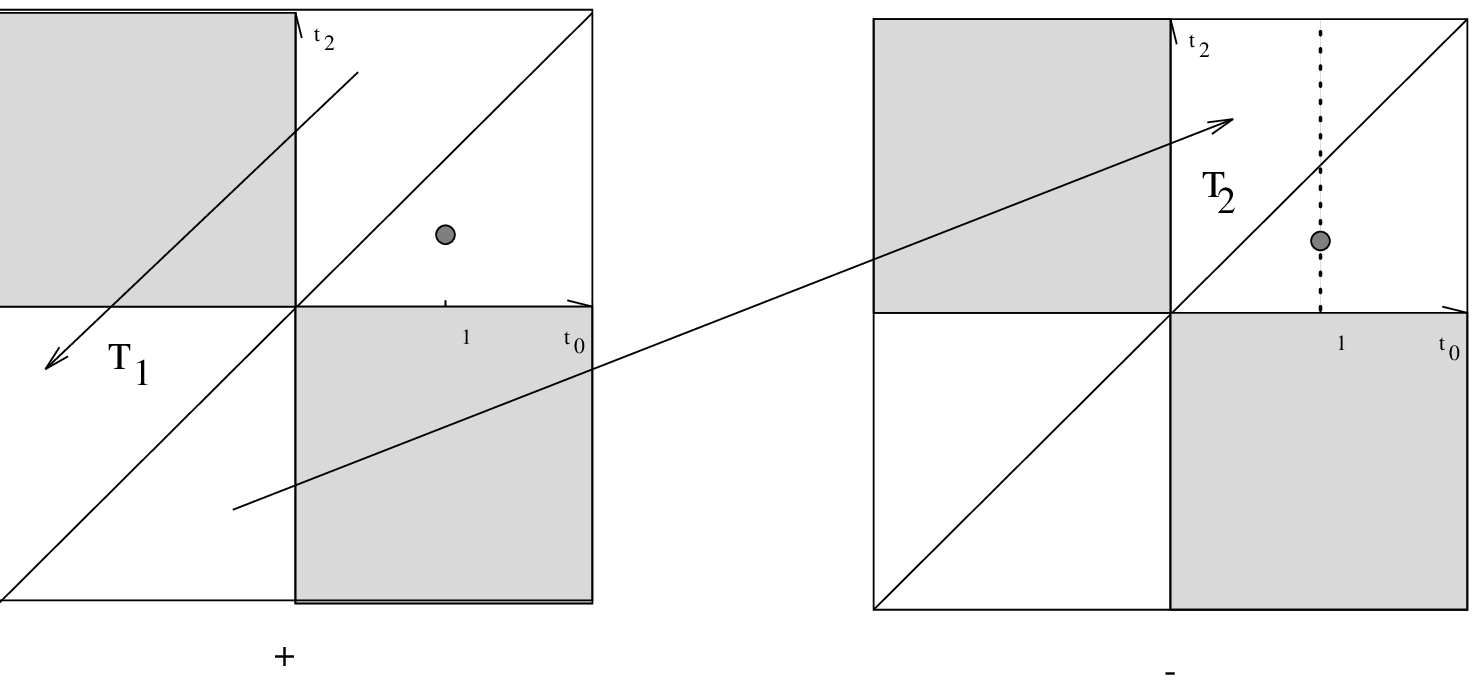}
\end{center}
\caption{\label{fig:param}}
\setlength{\baselineskip}{13pt}
Range of parameters $t_0$, $t_2$ for which the model (\ref{eq:Hamil}) is
integrable as a consequence of (\ref{eq:solvpt}) for $U=+2(t_0-t_2)$
(left) and $U=-2(t_0-t_2)$ (right). The dots mark the model
introduced in Ref.~\cite{karn:94}.
\end{figure}

\begin{figure}
\begin{center}
	\leavevmode
	\epsfxsize=1.00\textwidth
	\epsfbox{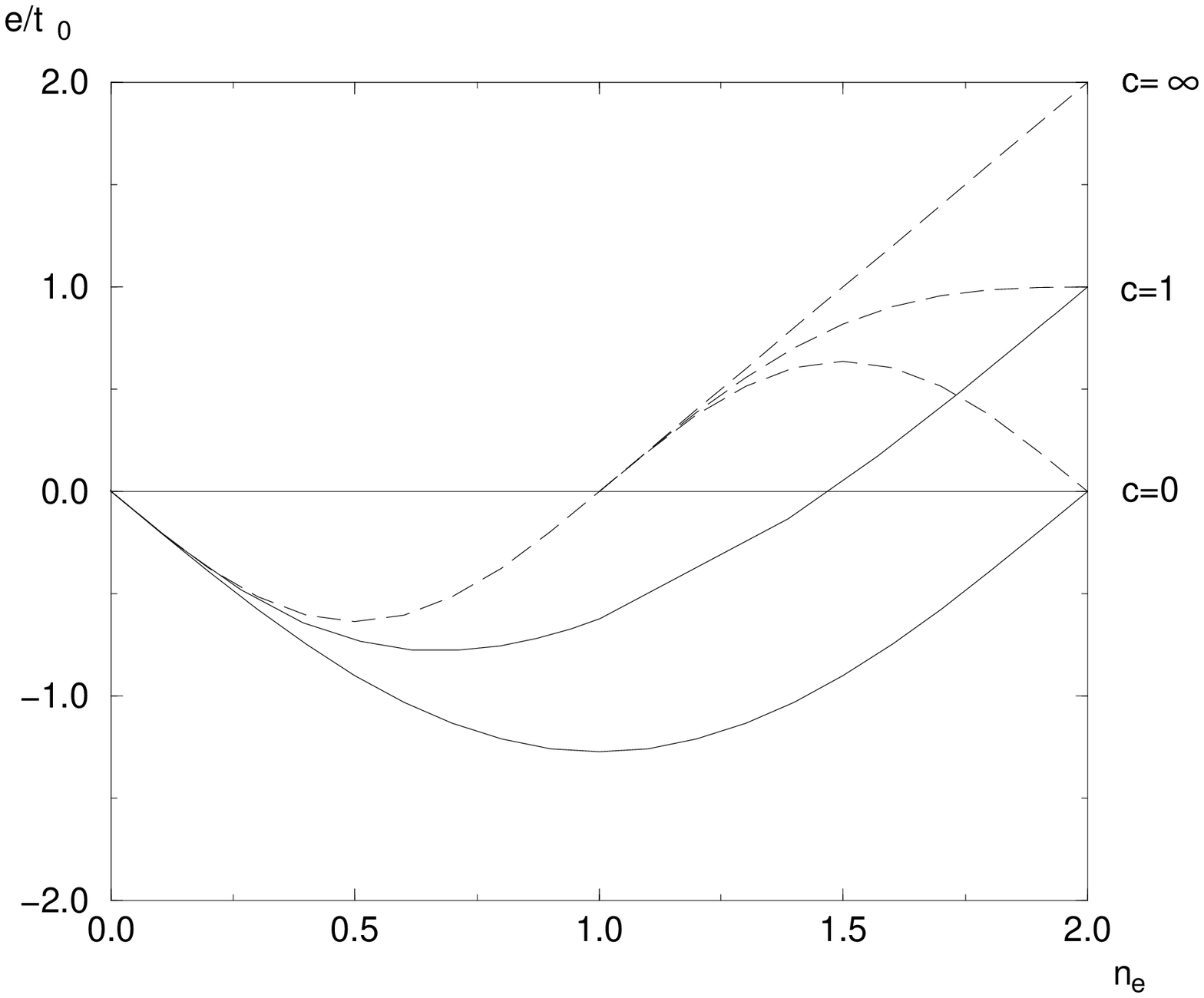}
\end{center}
\caption{\label{fig:gsrep}}
\setlength{\baselineskip}{13pt}
Energy of the antiferromagnetic ground state of the system
(\ref{eq:Hamil}) vs.\ electron density in the repulsive regime for
various values of the reduced coupling constant $c$. For comparison,
the energy of the ferromagnetic state (\ref{eq:E0-FM}) is also
included. Note, that for $c\to\infty$ the ferro- and antiferromagnetic
states are degenerate.
\end{figure}

\begin{figure}
\begin{center}
	\leavevmode
	\epsfxsize=1.00\textwidth
	\epsfbox{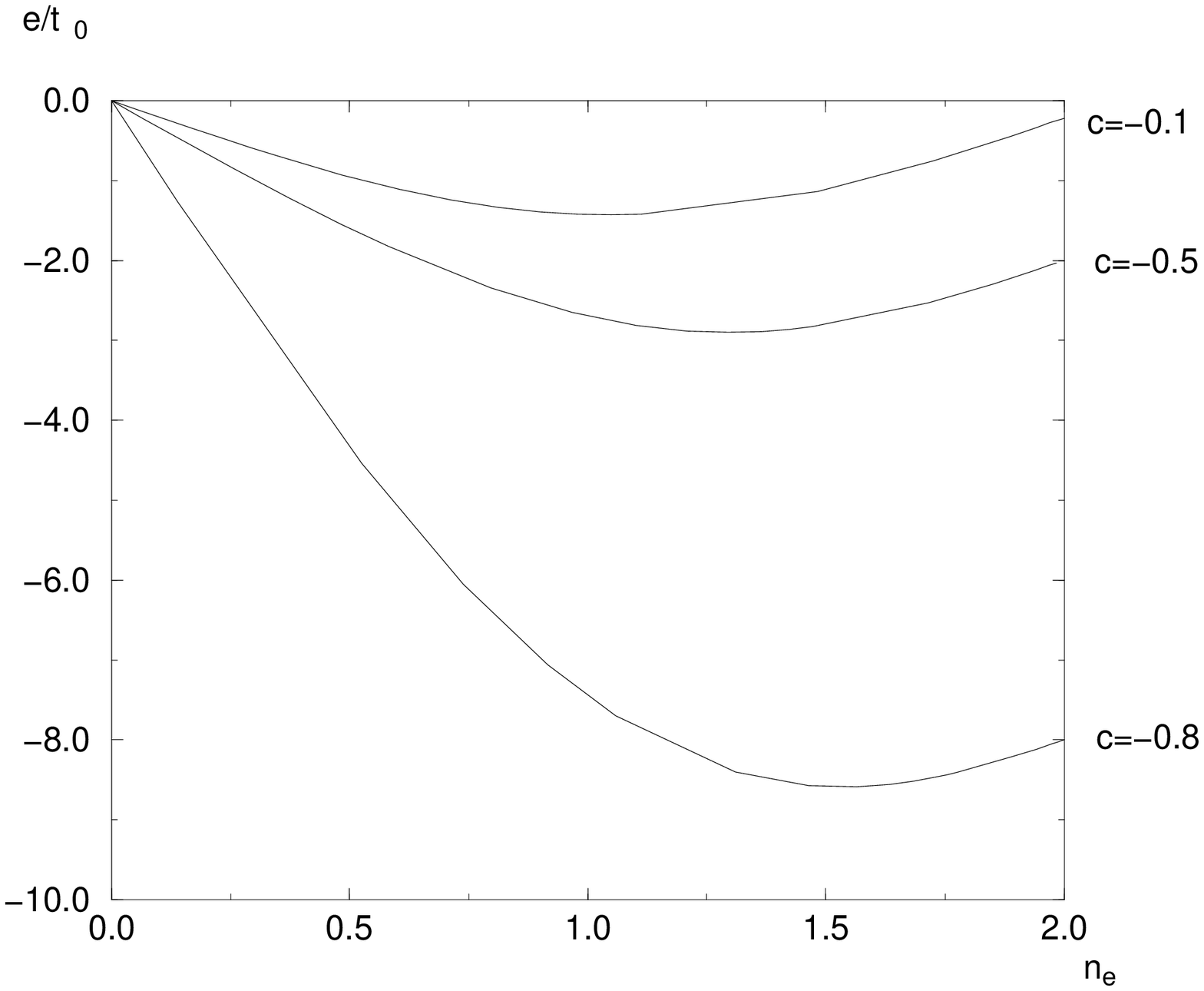}
\end{center}
\caption{\label{fig:gsatt}}
\setlength{\baselineskip}{13pt}
Energy of the antiferromagnetic ground state of the system
(\ref{eq:Hamil}) vs.\ electron density in the attractive regime for
various values of the reduced coupling constant $c$.~
\end{figure}

\begin{figure}
\begin{center}
	\leavevmode
	\epsfxsize=1.00\textwidth
	\epsfbox{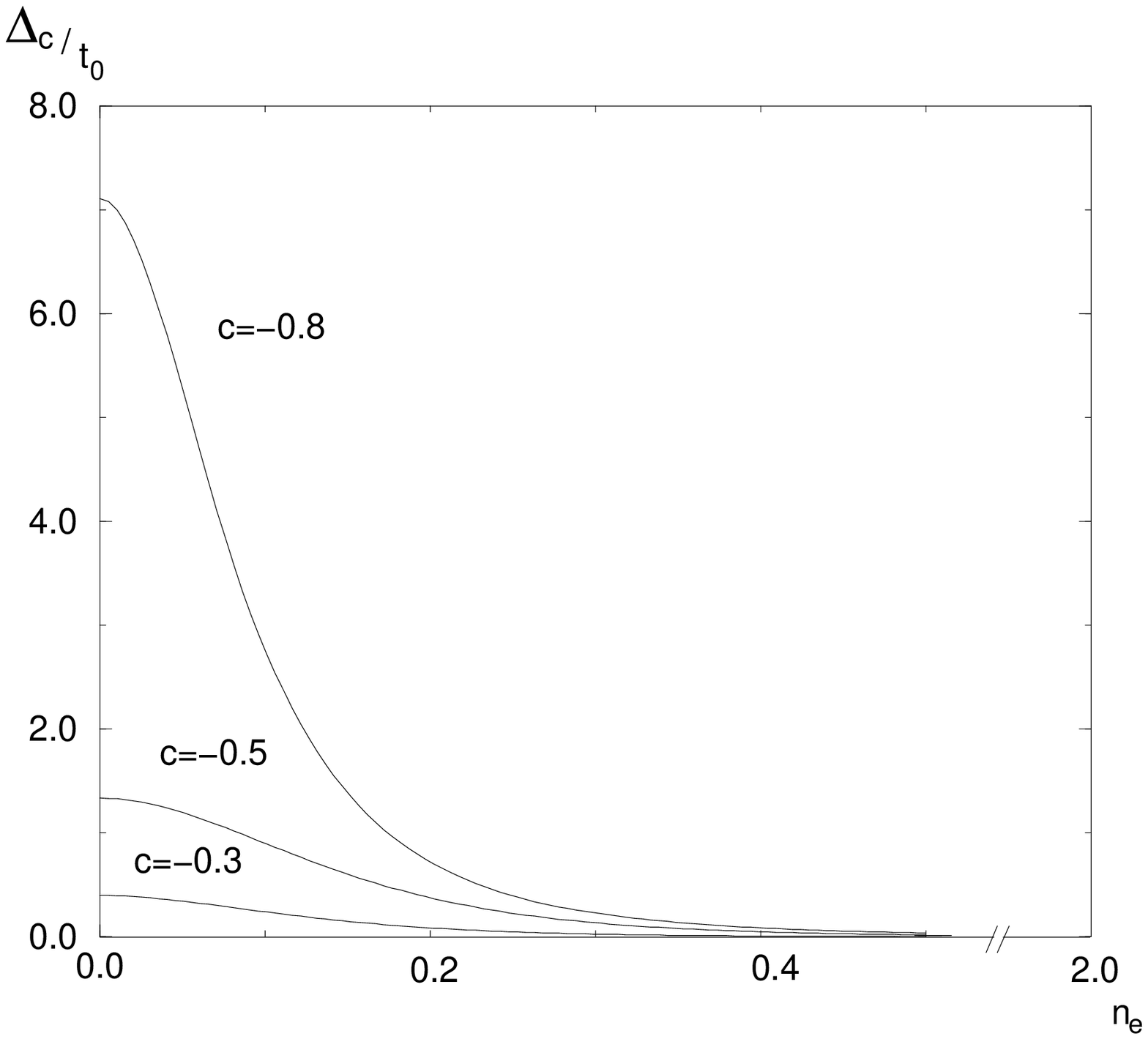}
\end{center}
\caption{\label{fig:gap}}
\setlength{\baselineskip}{13pt}
Energy gap for the creation of unpaired electrons as a functions of the
density of particles for several values of the parameter $c$.
\end{figure}

\begin{figure}
\begin{center}
	\leavevmode
	\epsfxsize=1.00\textwidth
	\epsfbox{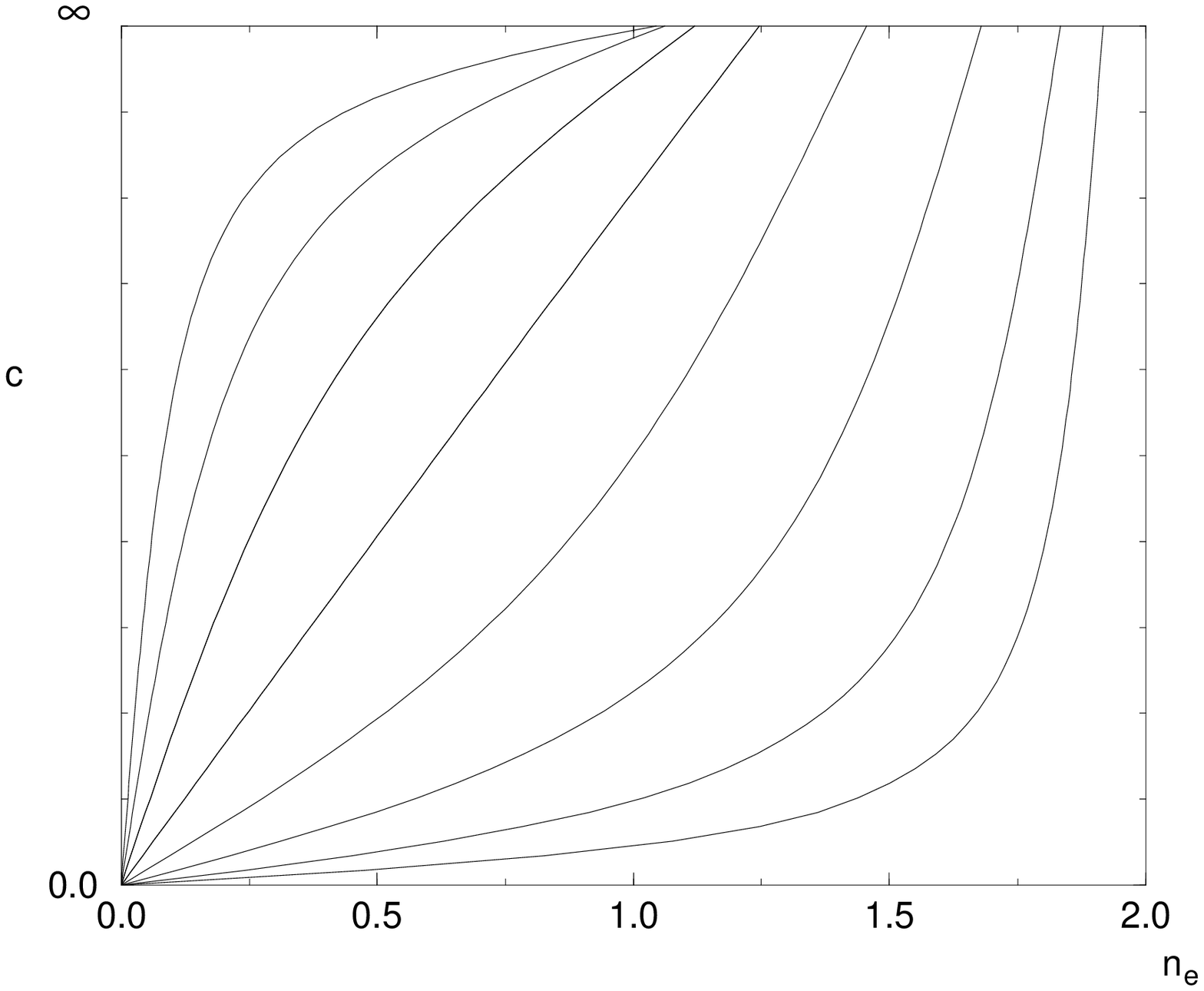}
\end{center}
\caption{\label{fig:dcrep}}
\setlength{\baselineskip}{13pt}
Contours of constant $\xi_c(Q)$ (and hence identical critical exponents)
in the $n_e$--$c$ parameter plane of the repulsive model. $\xi_c(Q)$
varies between $1$ (at low densities) and $\sqrt{2}$ (the free fermionic
case) for finite $c$.
\end{figure}

\begin{figure}
\begin{center}
	\leavevmode
	\epsfxsize=1.00\textwidth
	\epsfbox{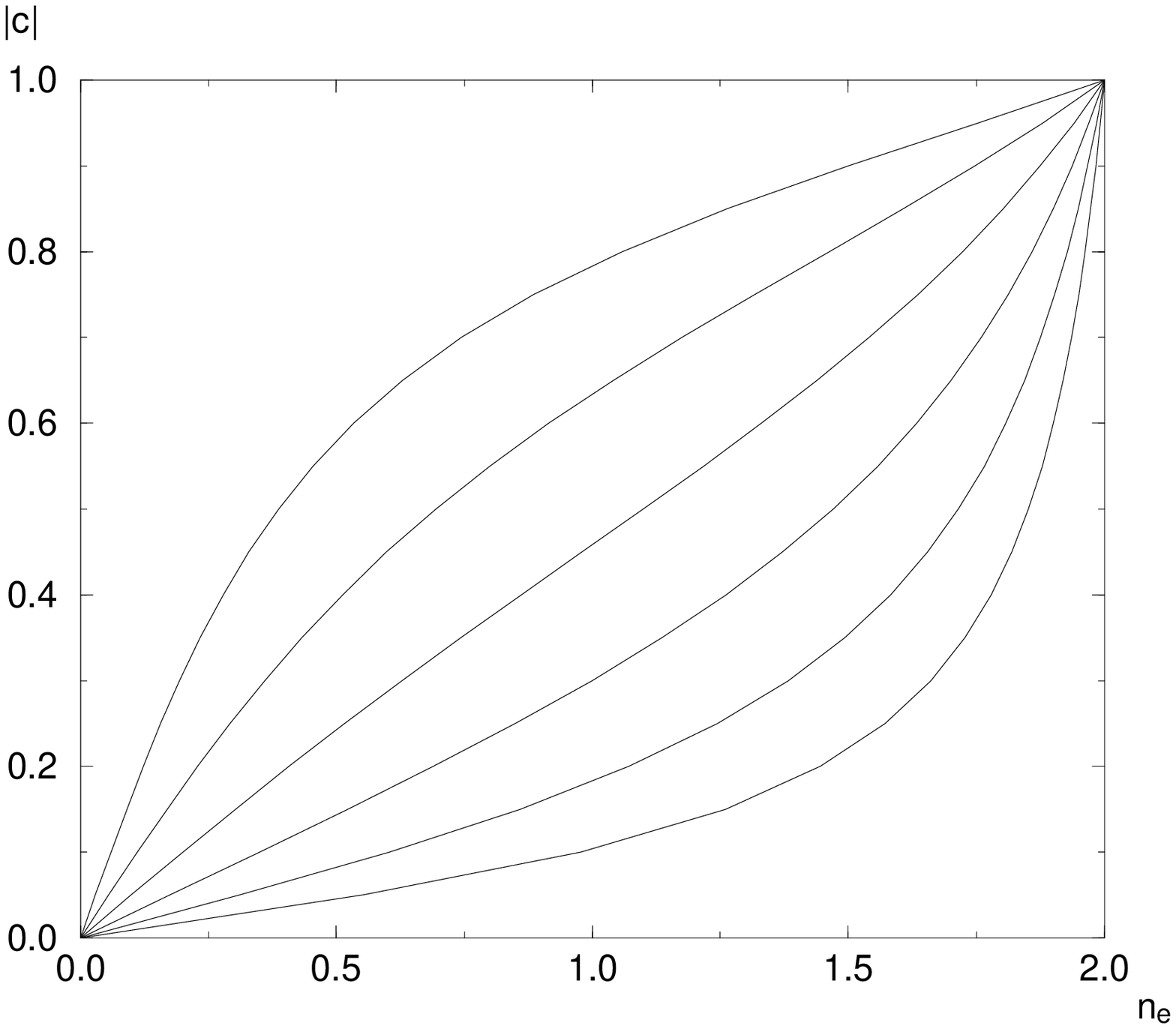}
\end{center}
\caption{\label{fig:dcatt}}
\setlength{\baselineskip}{13pt}
Contours of constant $\xi_p(Q)$ (and hence identical critical exponents)
in the $n_e$--$|c|$ parameter plane of the attractive model at small
magnetic fields $h<h_{c1}$. $\xi_p(Q)$ varies between $1$ and $1/\sqrt{2}$
for any finite $c$.
\end{figure}

\end{document}